\documentclass[amssymb,aps,twocolumn,floats,showpacs]{revtex4-1}
\usepackage{color,graphicx}
\usepackage{natbib}
\newcommand{\YK}{120$^{\circ}~$}
\newcommand{\J}{J_{AF}}
\newcommand{\Fig}{Fig.~}

\begin{document}

\title{Electronic route to stabilize nanoscale spin textures in itinerant frustrated magnets}

\author{Sahinur Reja$^1$, Jeroen van den Brink$^1$ and Sanjeev Kumar$^2$}
 
\address{
$^1$ Institute for Theoretical Solid State Physics, IFW Dresden, 01171 Dresden, Germany \\
$^2$ Indian Institute of Science Education and Research (IISER) Mohali, S.A.S. Nagar, Sector 81, Manauli PO 140306, India
}

\begin{abstract}
We unveil novel spin textures in an itinerant fermion model on a frustrated triangular lattice in the limit of low electronic density. Using hybrid Monte Carlo simulations on finite clusters
we identify two type of nanoscale spin textures in the background of \YK order
: (i) a planar ferromagnetic cluster, and (ii) and a non-coplanar cluster with spins oriented perpendicular to the \YK plane. Both these textures lead to localization of
the electronic wavefunctions and are in-turn stabilized by the concomitant charge modulations. The non-coplanar spin texture is accompanied by 
an unusual scalar chirality pattern. A well defined electric charge and magnetic moment associated with these textures allow for their easy manipulation by external
electric and magnetic fields -- a desirable feature for data storage.
We identify a localization-delocalization behavior for electronic wavefunctions which is unique to frustrated magnets, and propose a general framework for stabilizing similar spin textures in spin-charge coupled systems.
\end{abstract}

\date{\today} 


\maketitle

{\it Introduction ---} Extended magnetic objects that are large enough to be stable against thermal and quantum fluctuations are in common use to densely store information. In materials with immobile magnetic bits, such as for instance patches of magnetization in hard ferromagnets, there are essentially two ways to address individual bits.
If one wishes steer clear of the overhead of wiring each bit individually, the choice that remains is to physically move the storage medium with respect to the reading/writing device. In materials with {\it mobile} magnetic information carriers the required mechanically moving parts can in principle be avoided by propagating instead the mesoscopic magnetic bits themselves, for instance by applying an electric field in a race-track memory set-up~\cite{Parkin2008,Bauer2013}. 
%
% Science 11 April 2008:  Vol. 320 no. 5873 pp. 190-194  DOI: 10.1126/science.1145799, Magnetic Domain-Wall Racetrack Memory, Stuart S. P. Parkin, Masamitsu Hayashi, Luc Thomas
%
In the past years this concept has given a strong impetus to theoretical and experimental research on mobile magnetic textures, such as single magnetic skyrmions, polarons, bubbles and domain walls~\cite{Jiang2015,Romming2013,Nagaosa2013,Seki2012,Muhlbauer2009,Rossler2006,Yu2010,Moon2015,Dietl2010}. 
On the theoretical side, the focus so far has been on finding novel spin textures in purely magnetic systems. The candidate models are the anisotropic Heisenberg models in the 
presence of external magnetic field~\cite{Muhlbauer2009,Rossler2006,Okubo2012,Leonov2015}.

In this Letter, we show that such novel spin textures resembling magnetic polarons can emerge in frustrated antiferromagets made out of spins and charges on a triangular lattice, which in their phase diagram are close to a region of phase separation. The results are obtained via unbiased Monte Carlo simulations on finite lattices, and crosschecked with variational calculations.
We find that ferromagnetic polarons with two different spin orientations are stabilized as a consequence of competing interactions on the frustrated lattice geometry. These spin textures derive their stability from the fact that they appear together with a charge modulation. As opposed to most other magnetic skyrmion and domain-wall textures these spin-textures are thus heavily charged and mobile. This implies that they can easily be manipulated with applied electric and magnetic fields, which is a useful feature for data storage. From a fundamental standpoint these results provide a connection between localization physics, frustrated magnetism and novel spin textures. From the charge profiles we infer the presence of an interesting localization to delocalization behavior for electronic wavefunctions that is specific to frustrated lattices.

{\it Model and Method --} The Hamiltonian that we consider consists of localized classical moments Heisenberg-coupled to nearest neighbors and Kondo-coupled to itinerant fermions residing on a triangular lattice. The interplay of conduction electrons and the frustrated geometry give rise to many interesting effects, 
such as for instance a multiple-Q magnetic order~\cite{Martin2008, Kumar2010, Akagi2010a, Chern2010, Hayami2014}, the anomalous Hall effect~\cite{Martin2008,Akagi2010a,Ishizuka2013b,Chern2014}, coupled spin-charge phases~\cite{Misawa2013,Reja2015,Akagi2015}, partially disordered phases~\cite{Ishizuka2012a,Ishizuka2013a}. We work in the limit of strong Kondo coupling and focus explicitly on the low electronic density limit. 
In the strong coupling limit the electronic spin gets slaved to the local moment and the Kondo model reduces to a double-exchange (DE) model. The resulting Hamiltonian for spinless fermions is given by,
\begin{eqnarray}
H &=& -\sum_{ \langle ij \rangle}
t_{ij} \left ( c^{\dagger}_{i} c^{~}_{j} + H.c. \right ) 
+ J_{AF} \sum_{ \langle ij \rangle} {\bf S}_{i} \cdot {\bf S}_{j},
\end{eqnarray}
where $c^{}_{i}$ ($c^{\dagger}_{i}$) is the electron annihilation (creation) 
operator with spin parallel to the local magnetic moment ${\bf S}_i$. 
The angular brackets in the summations denote the nn pairs of sites on a triangular lattice. 
$J_{AF}$ is the strength of the AF coupling between the localized spins.
The projected hopping is given by,
$t_{ij} = t_0 [\cos(\theta_i/2)\cos(\theta_j/2) + \sin(\theta_i/2)\sin(\theta_j/2) e^{{\rm -i} (\phi_i - \phi_j)}]$, where $\theta_i$ ($\phi_i$) is the polar (azimuthal) angle
for localized spins at site $i$, and
$t_0$ is the bare fermionic hopping amplitude between nn sites. Here onwards we set $t_0 = 1$ as the unit of energy, and therefore the only free parameter in the model
is $J_{AF}$.
The model has been extensively studied in the context of colossal magnetoresistance \cite{Dagotto2001, kumar2006insulator, Sen2007}, and more recently for frustrated itinerant magnets \cite{Kumar2010, Venderbos2012, Barros2014, Ishizuka2015}.

Here we focus on the limit of very low electronic filling in order to understand the effect of itinerant electrons on a frustrated magnetic state.
Note that the DE model alone ($J_{AF} = 0$) leads to a ferromagnetic ground state even for a single itinerant electron. On the other hand, in the absence of electrons
the ground state supports a \YK order for all non-zero values of $J_{AF}$.
While the overall ground state phase diagrams for this model suggest a tendency towards electronic phase separation in the low-density regime, 
the detailed real-space nature of the states is not known \cite{Akagi2011}.
In particular, one can ask if certain special spin textures are induced as a consequence of a single electron trying to gain kinetic energy in an otherwise frustrated magnetic structure. 
In order to find the answer, we make use 
of a hybrid Monte Carlo method which combines the classical Monte Carlo for spins with numerical diagonalization for fermions~\cite{Yunoki1998}. The method is numerically exact, and 
has served as a very useful tool for exploring the physics of spin-charge coupled systems~\cite{suppl}. 
The simulations begin at high temperature with a random spin configuration, and the the temperature is then decreased in small steps \cite{suppl}.
The results presented here correspond to low temperatures, $k_BT \sim 0.002 t_0$.

\begin{figure}
\centerline{
\includegraphics[width =.7\columnwidth,angle=-90]{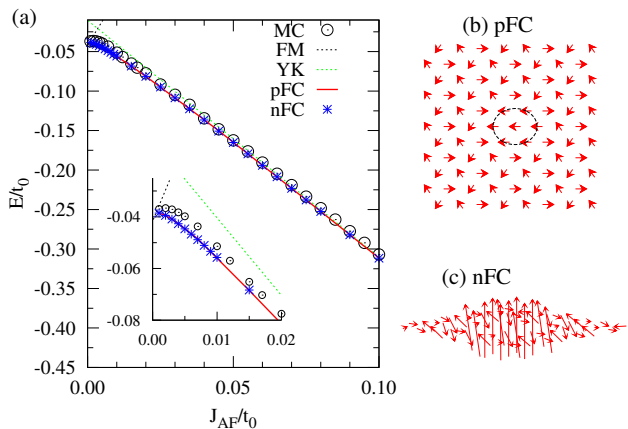}
}
\caption{(Color online) (a) Total energy per site for different values of $J_{AF}$. The empty circles are the Monte Carlo results, the solid line is obtained from variational 
calculations for planar FM cluster (see text), and the star symbols are restricted Monte Carlo simulations where some sites were forced to retain the $120^{\circ}$ order.
The dashed line with positive (negative) slope represents the energy of a long-range ordered FM (\YK) state. Two typical Monte Carlo snapshots showing, (b) a planar FM cluster and
(c) A non-coplanar FM cluster, residing in the \YK background.}

\label{phases_one_el}
\end{figure}

\begin{figure}
\begin{center}
\vspace{-1cm}
\includegraphics[width =0.9\columnwidth,angle=0]{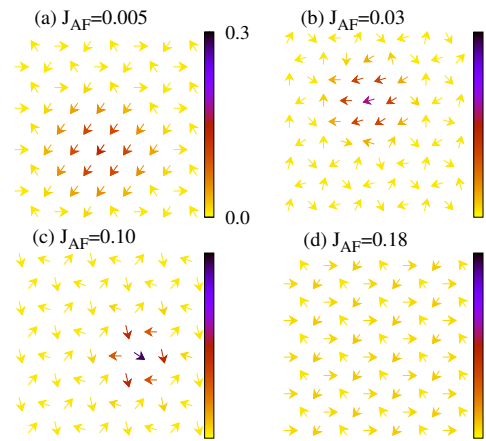} 
\end{center}
\vspace{0.1cm}
\caption{(Color online) 
(a)-(d) Spin and charge configurations at different $J_{AF}$ values as obtained from MC simulations where spins were restricted to remain in a plane. 
The color of the spin represents the charge density at that site. The charge tends to initially localize and then delocalize upon increasing $J_{AF}$ values.
}
\label{fm_spin_charge_config}
\end{figure}

{\it Results and Discussions:}
The total energy per site for the case of one electron on a 12$\times$12 lattice is
shown in \Fig\ref{phases_one_el}(a). As expected, the MC energy (circles) interpolates between the FM and the \YK energy asymptotes which are plotted as dashed lines. 
A careful analysis of the low-temperature real-space spin configurations indicated the presence of two interesting 
spin-textures in the intermediate $J_{AF}$ regime, $0.01 \leq J_{AF} \leq 0.05$. The first one is a planar FM cluster (pFC) and the second 
is a non-coplanar FM cluster (nFC) with additional meron-like texture (see \Fig\ref{phases_one_el}(b)-(c)). These two spin textures are almost degenerate in energy over the 
entire intermediate $J_{AF}$ range.
Note that it is difficult to identify the detailed nature of such textures from a calculation of spin structure factors, as the majority of the background retains the 
\YK order and therefore the structure factor will be dominated by the \YK state. Taking hints from the textures obtained in the Monte Carlo simulations, we set-up restricted Monte Carlo simulations 
in order to better understand the nature of these spin patterns. 

\begin{figure}
\centerline{\includegraphics[width =.7\columnwidth,angle=-90]{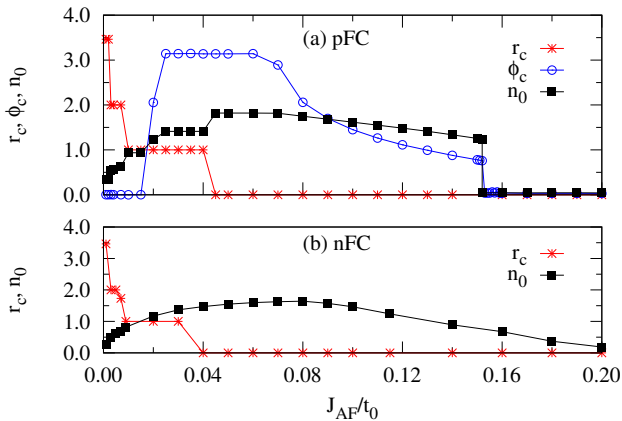}}
\caption{(Color online) The radius (red stars), $r_c$, and the angle relative to the $120^{\circ}$ state (blue circles), $\phi_c$, of the planar FC as a function of 
the AF coupling strength $J_{AF}$, as obtained from Monte Carlo simulations where spins are restricted to be in a single plane. The filled squares represent the charge density, $n_0$, at the 
central site of the planar FC. The charge density is scaled up by a factor of $5$ in order to clarify the variations.
(b) The variation of the radius, $r_c$, of the non-coplanar FC and $n_0$ inferred from a second set of restricted Monte Carlo simulation (see text).}

\label{cluster_size_phi}
\end{figure}

In the first scheme we restrict the spins to lie in a single plane so that the spins are
effectively XY type. This can also be thought of as a model for easy plane triangular magnet. The ground state spin and charge configurations obtained from
these simulations are shown in \Fig\ref{fm_spin_charge_config}. While the large regions of the lattice still follow the \YK pattern, a cluster of aligned spins emerges spontaneously.
The charge density is large over these aligned spins, and is essentially zero elsewhere. This provides a simple picture of electronic self-trapping in the \YK background: a single electron 
tries to make a FM cluster so that it can move freely over this cluster, however there is a price to be paid in terms of the $J_{AF}$. Therefore, one expects that the size of the FM cluster 
should decrease upon increasing $J_{AF}$. Indeed, this is reflected in the real-space plots shown in \Fig\ref{fm_spin_charge_config}. However, even though the size of the FM cluster decreases
monotonically, the charge patterns indicate a non-monotonic localization-delocalization crossover. To investigate further the variations of FM cluster size and the charge localization, we 
perform variational calculations by allowing for FM clusters with different radii, and different orientations w.r.t. the \YK order. This is achieved by orienting all spins inside a ring of 
radius $r$ in same direction, while the remaining spins retain the \YK pattern. For each value of $J_{AF}$ the FM cluster is varied in radius as well as orientation to obtain the 
minimum energy. The results are plotted in \Fig\ref{cluster_size_phi}(a). While the radius of the FM cluster decreases monotonically with increasing $J_{AF}$, the orientation shows a non-monotonic
behavior. $r_c \rightarrow \infty$, $\phi_c = 0$ correspond to a FM phase, and $r_c = 0$, $\phi_c = 0$ is the \YK phase. Thus the variational calculation captures naturally the two limiting phases. 
In the same plot we show the charge at the central site, $n_0$, of the FM cluster, as a function of $J_{AF}$. There is a non-monotonic variation in $n_0$ which indicated an 
interesting localization-delocalization crossover. While the localization is gradual with charge at central site increasing to its maximum possible value around $J_{AF} \sim 0.05$, the
delocalization is abrupt near $J_{AF} \sim 0.15$.

\begin{figure}
\begin{center}
%\vspace{-.6cm}
\includegraphics[width =.9\columnwidth,angle=0]{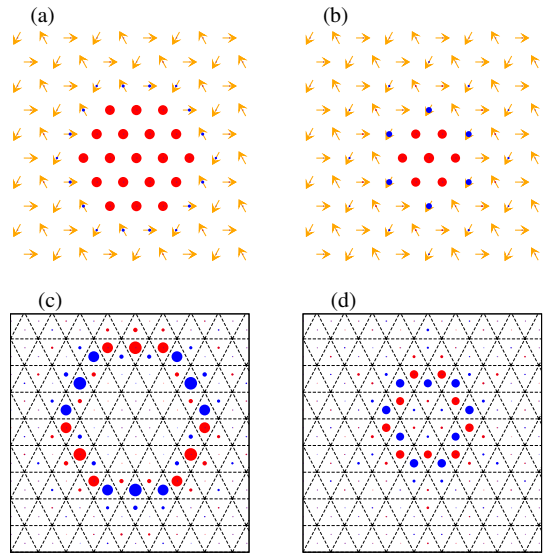} 
\end{center}
\vspace{0.1cm}
\caption{(Color online) (a)-(b) The non-coplaner spin configuration at $J_{AF} = 0.005$ and $J_{AF} = 0.03$ obtained within restricted MC simulations, where the boundary layers are forced to
retain \YK order and a central spin in fixed to be normal to the \YK plane.
The magnitude (sign) of $S_z(i)$ is shown as the size (color) of the circles. (c)-(d) The scalar chirality pattern for the 
two values of $J_{AF}$.}
\label{spin_config_ring}
\end{figure}

As mentioned earlier, another spin texture is almost degenerate in energy with the pFC. This is a non-coplanar spin arrangement where 
a cluster of spins orients normal to the \YK plane. 
To analyze this structure and its stability further we perform another set of restricted Monte Carlo simulations. 
We force the boundary layer of spins to retain the \YK order, and a central spin to point perpendicular to 
the \YK plane. The remaining spins are updated using the Monte Carlo simulations \cite{suppl}.
The resulting ground state spin configurations are shown in \Fig\ref{spin_config_ring} (a)-(b) for two values of $J_{AF}$.
Here FM clusters point perpendicular to \YK spin configuration. On the boundary
of the FM cluster, $z$-components of the spins on different rings oscillate in sign before they vanish approaching the boundary. 
The $z$-components of all spins at each site are also shown in panels (a) and (b). The size of the filled circles represent the values of $z$-components and 
colors (red or blue) represents their sign. The corresponding charge distributions (not shown) are again correlated with the formation of FM clusters.
The radius of the non-coplanar FC, together with the density at central site $n_0$ are shown in \Fig\ref{cluster_size_phi}(b). 
Similar to the case of planar FC, the cluster size decreases monotonically upon increasing $J_{AF}$ and the variation in $n_0$ is non-monotonic
Note that the nFC texture is not simply a pFC texture rotated by $\pi/2$, but contains interesting modulations of the $z$-components of the spins
(see \Fig\ref{spin_config_ring}). In order to highlight the noncoplanar character of the nFC texture, we compute the associated chirality patterns.
These are shown in \Fig\ref{spin_config_ring} (c)-(d), where $\chi_{ijk} = {\bf S}_i \cdot ({\bf S}_j \times {\bf S}_k)$ for each triangle with vertices
$i$,$j$ and $k$ is shown in the center of the triangle. The size of the circle represents the magnitude of the chirality and the color indicates the sign. 
The positive and negative chiralities are intertwined to result in a zero net chirality. The textures represent examples of localized chirality quadrupoles.

\begin{figure}
\begin{center}
%\vspace{-.6cm}
\includegraphics[width =.95\columnwidth,angle=0]{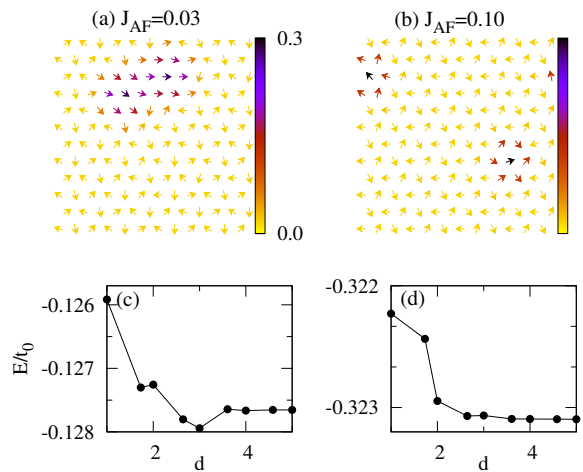} 
\end{center}
\vspace{0.1cm}
\caption{(Color online) The spin-charge textures for the case of two electrons
for (a) $J_{AF}=0.03$ and (b) $J_{AF}=0.1$. (c) Variation of energy with distance $d$ between the centers of the two $r_c = 1$ planar FM cluster for $J_{AF}=0.03$.
(d) Same as (c) for $J_{AF} = 0.1$, where only one spin is deviated from the \YK order.} 
\label{two_elec}
\end{figure}

In order to test the stability of these textures in a realistic scenario, it is important to look beyond the single electron case. 
While a genuine finite-density system would require 
us to go to larger lattices which is not feasible within the exact simulation scheme, we present results for two electrons on a 12$\times$12 lattice \cite{suppl}.
\Fig\ref{two_elec}(a) and (b) display the ground state spin-charge configuration obtained from the MC simulations with XY spins with two electrons for $\J=0.03$ and $\J=0.10$, respectively.
The spin-charge textures obtained in the one-electron case retain their identity and their separation is determined by the value of $J_{AF}$. It is important to note that two electrons do
not lead to a larger FM cluster with uniform charge density.
For a better understanding of the two-electron results in terms of single electrons, we study the effective interaction between two pFC textures.
This is achieved by assuming a fixed radius for the two pFCs and plotting the total energy as a function of their separation $d$. The resulting 
plots are shown in \Fig\ref{two_elec}(c) and (d) for the two representative values of $J_{AF}$. For $J_{AF} = 0.03$, we find that the energy is
minimum at a separation of $d=3$. Indeed, the separation found in the Monte Carlo simulations is close to $3$ lattice spacings. For $J_{AF} = 0.10$,
the energy keeps decreasing monotonically with increasing $d$, indicating an effective repulsive interaction between two pFC textures. The Monte 
Carlo results are consistent with this simple picture and the two textures reside away from each other (see \Fig\ref{two_elec}(b)).

A remark is in order concerning the generality of such spin-charge textures in Kondo-lattice Hamiltonians. A thermodynamic limit
argument suggests that for any finite electronic density inside the phase separation regime, the system will show macroscopic phase coexistence.
However, this assumes that the boundary contributions are negligible compared to bulk contributions. This assumption does not hold if the lattice is not infinite
or if the volume fraction of one of the phases is much smaller. Therefore, such spin textures can be generally expected if the density
lies close to one of the phase separation boundaries in mesoscopic systems. 

{\it Conclusion:}
We show that electronic Hamiltonians involving coupled spin and charge degrees of freedom are capable of stabilizing novel spin-charge textures in the ground state.
We study a prototype model for coupled spin and charge degrees of freedom using a numerically exact Monte Carlo simulation method.
In the low-density limit, we find that two novel spin-charge textures are stabilized as a consequence of competing interactions on a frustrated lattice.
These are the planar FM cluster and the non-coplanar FM clusters, both of which exist together with a localized charge-density profile. 
We also propose a general scheme for stabilizing such
spin textures in electronic Hamiltonians involving coupled spin and charge variables.
The search for stable and mobile spin textures in magnetic systems is a rapidly evolving research direction due to
the potential applications in data storage devices with easy read/write operations.
In contrast to the conventional mechanism of stabilizing
novel spin textures in magnetic models, which involves anisotropic interactions, longer-range interactions and external magnetic fields~\cite{Muhlbauer2009,Rossler2006,Okubo2012,Leonov2015}, 
the electronic route presented in this letter can open up new possibilities.

{\it Acknowledgments:} This work is supported by SFB 1143 of the Deutsche Forschungsgemeinschaft. SK acknowledges hospitality at IFW Dresden during June-July 2015, and support from the Department of Science and Technology (DST), India.

%\bibliographystyle{apsrev4-1} %{apsrev}
%\bibliography{Frus_triangular,few_el_triangle,My_Papers}

%\include{manuscript_v1.bbl}

%merlin.mbs apsrev4-1.bst 2010-07-25 4.21a (PWD, AO, DPC) hacked
%Control: key (0)
%Control: author (72) initials jnrlst
%Control: editor formatted (1) identically to author
%Control: production of article title (-1) disabled
%Control: page (0) single
%Control: year (1) truncated
%Control: production of eprint (0) enabled
%

\end{document}

\title{Supplementary Material for\\ ``Electronic route to stabilize nanoscale spin textures in itinerant frustrated magnets''}

\author{Sahinur Reja$^1$, Jeroen van den Brink$^1$ and Sanjeev Kumar$^2$}

\address{
$^1$ Institute for Theoretical Solid State Physics, IFW Dresden, 01171 Dresden, Germany \\
$^2$ Indian Institute of Science Education and Research (IISER) Mohali, Sector 81, S.A.S. Nagar, Manauli PO 140306, India \\
}

\date{\today}
\begin{abstract}
      In this supplement we include the following material: (i) the details of the hybrid Monte Carlo method and the restricted simulation schemes used in this work, 
      (ii) results on the effect of finite lattice sizes and that of increasing number of electrons on the stability of the spin textures obtained in our study.
\end{abstract}

\maketitle

\subsection{Hybrid Monte Carlo Method}
 The model Eq. (1) of the main text is investigated using a hybrid Monte Carlo (MC) method which combines
 the classical Monte Carlo for spins with the fermion diagonalization. The solution of a new Schroedinger equation is required at each Monte Carlo update step in order to obtain the 
 electronic contribution to the total energy of a given classical configurations of localized spins.
 This is achieved by numerical diagonalization of the Hamiltonian matrix at each Monte Carlo step.
 The standard Metropolis algorithm is then followed for spin updates. The computational time with number of lattice sites scales as 
 $N^4$ ($N$ from Monte Carlo and $N^3$ from matrix diagonalization). 
 Most of the results presented in this work are obtained on clusters with $N=12^2$ sites. 
 Typically $\sim 10^4$ MC steps are used for equilibration at each temperature, and a similar number of steps for 
 computing thermal averages over classical spin variables. We begin the simulations with a random spin
 configuration and at a high temperature ($k_BT \sim 0.3$) and reduce stepwise in about $30$ temperature steps to very low temperature ($k_BT \sim 0.002$). 
 The results at these low temperatures are taken as representative of the ground state properties.
 
 Since the focus of our work is the formation of nanoscale textures in the low density limit, the usual thermodynamic 
 quantities such as the structure factors, etc. are not very useful. Since most of the region retains a \YK order, the structure factor will only
 reflect the presence of that order.
 Therefore, the identification of textures
 requires a careful real-space analysis. In order to achieve this we have made use of restricted Monte Carlo simulations,
 where the type of restrictions we impose on spin variables is motivated by the structures obtained from fully unbiased simulations.
 From the full simulations we could identify the tendency for formation of two type of textures, in the background of the \YK order: (i)
 the planar ferromagnetic cluster, where the spins of the cluster align along a common direction within the \YK plane, and (ii) an out of plane ferromagnetic
 cluster with modulations in $S_z$, which has a cluster of spins lifting out of \YK plane. We used two different restricted Monte Carlo schemes
 in order to further check the energetics of these spin textures.
 
\begin{figure}[t!]
\begin{center}
\includegraphics[width =0.75 \columnwidth,angle=270]{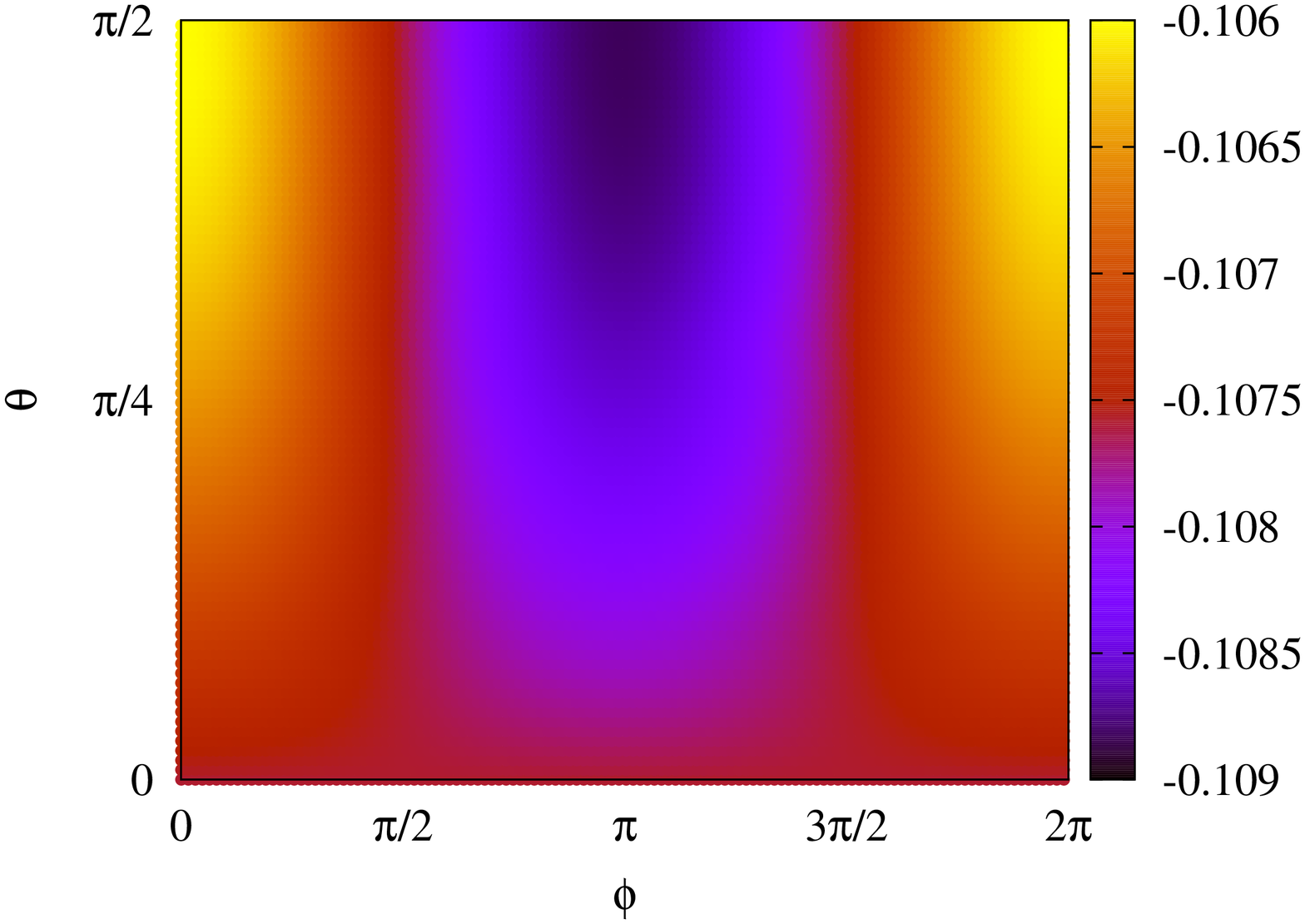}
\end{center}
\caption{Color-map plot of total energy as a function of polar and azimuthal angle of the $7$-site FM cluster. The minimum energy corresponds to $\theta = \pi/2$ and $\phi = \pi $. 
}
\label{spin_charge_diff_size}
\end{figure}

 \subsection{Restricted Monte Carlo Simulations}

 In order to check the existence and stability of the planer ferromagnetic cluster (pFC) in the \YK background, we perform MC simulation
 by limiting the spins to remain in XY plane. Indeed, the energy obtained in these planer simulations is further lowered compared to that obtained in the unrestricted
 simulations (see Fig. 1 in main text).

 As mentioned earlier, Monte Carlo simulations also indicate the stability of textures where a cluster of spins gets aligned and points perpendicular to the 
 \YK plane.
 In order to verify the stability of such a non-coplaner ferromagnetic cluster (nFC) we perform another set of restricted simulations.
 In these simulations, one of the spins is forced to point along $z$ direction and all the spins that are located at a distance of $r/a > 4$ ($a$ being the nearest neighbor distance)
 from the fixed spin are forced to retain the \YK order in $XY$ plane. 
 The choice of the cut-off of $r/a > 4$ is motivated by the fact that the unbiased calculations lead to a localization of charge density of the electron
 over sites that are well within the cut-off. Therefore, in the ground state \YK pattern is expected to be retained in these charge-free regions.
 The remaining spins are then updated using the standard Metropolis algorithm. These simulations also improve the total energy 
 of the system compared to unrestricted simulations. In fact, the value of total energy of configurations containing pFC and nFC textures lie very close to each other. 
 As discussed in the main text, the nFC texture also has
 modulations in the $z$ component of the spins, which seem to crucial for the stability of the texture from an energetic point of view.
 Therefore, it is important to emphasize that the nFC is not simply a pFC texture rotated by $\pi/2$. This is explicitly confirmed by the color-map plot of total energy shown in Fig. S1.
 The total energy of a $7$-site cluster is plotted as a function of polar angle $\theta$ and azimuthal angle $\phi$. 
 The minimum energy is obtained for one specific choice of the ($\theta$, $\phi$) pair, indicating an anisotropic nature of the pFC. This is in contrast to
 the standard magnetic polarons living in the paramegnetic background.

 %\vspace{0.8cm}
 
\subsection{Effect of Lattice Size and Electron Numbers}

\begin{figure}[t!]
\begin{center}
\includegraphics[width =0.96 \columnwidth,angle=0]{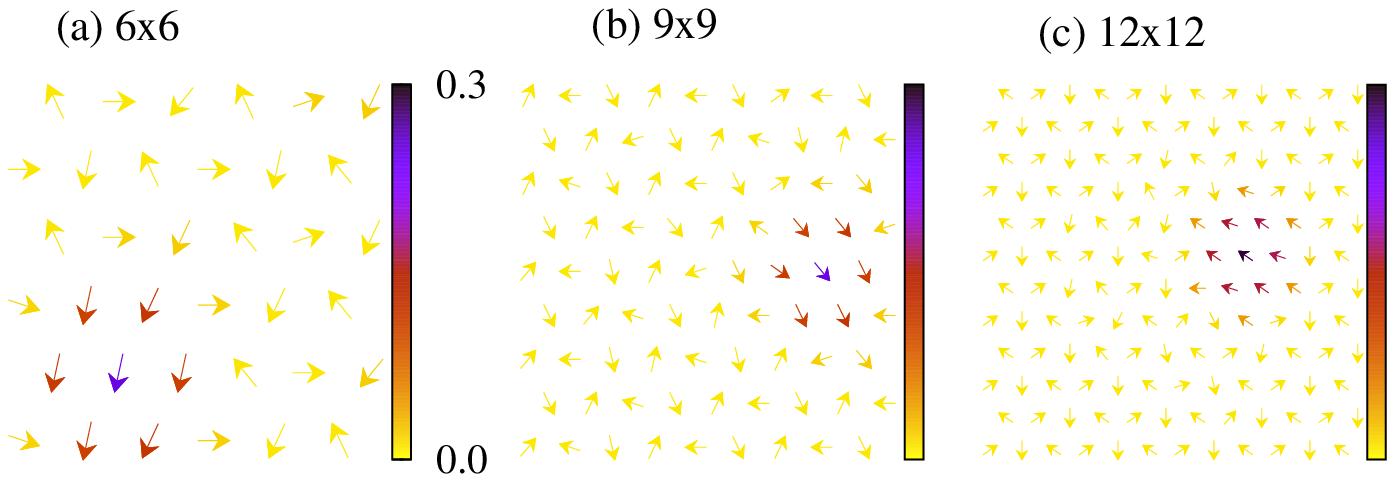}
\end{center}
\caption{ Spin and charge configurations at $J_{AF}=0.03$ with one electron on clusters with, (a) $6^2$,
 (b) $9^2$ and (c) $12^2$ sites respectively. The arrow directions indicate the orientation of planar spins and the color on the arrows represents the electronic charge density. 
}
\label{spin_charge_diff_size}
\end{figure}

\begin{figure}[t!]
\begin{center}
\includegraphics[width =0.96 \columnwidth,angle=0]{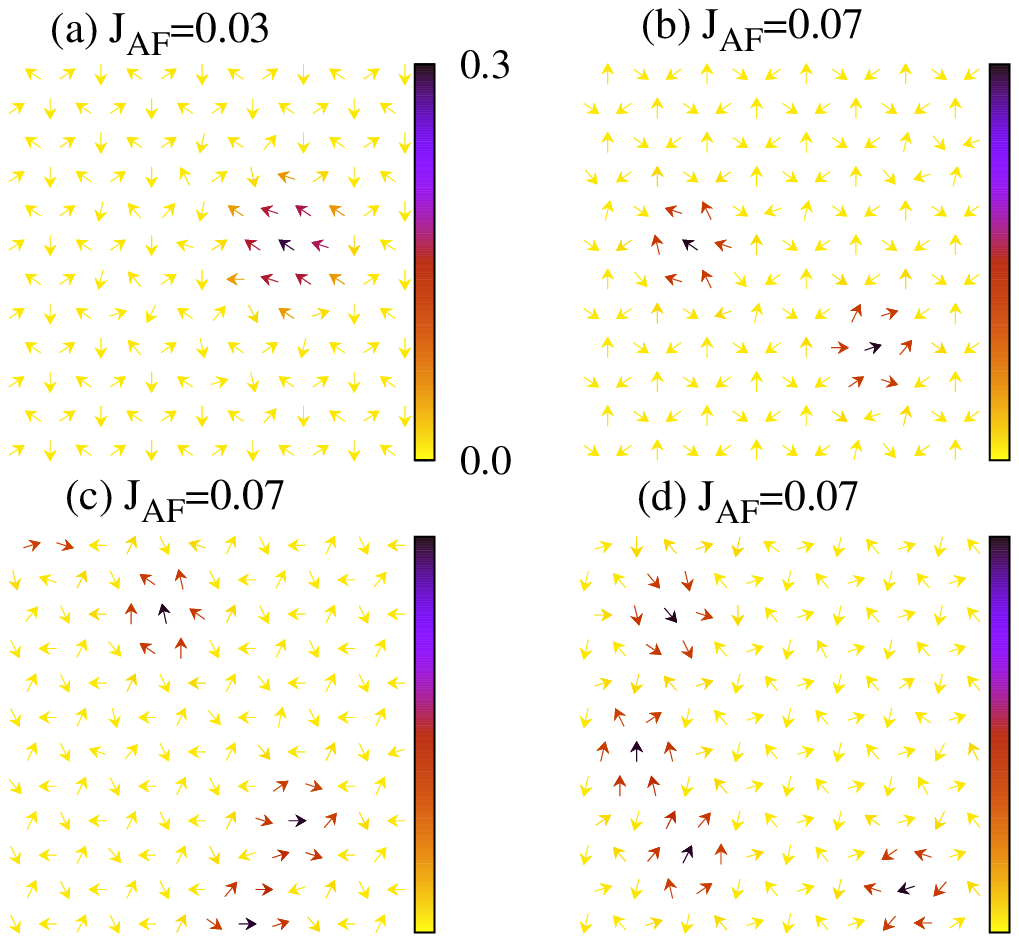}
\end{center}
\caption{(a)-(d) Spin and charge configurations on a lattice with $N = 12^2$
for one, two, three and four electrons respectively. The values of $J_{AF}$ are indicated in the plot. 
The arrow directions indicate the orientation of planar spins and the color on the arrows represents the electronic charge density.
Note that a larger FM cluster is not formed and these textures remain stable as individual dressed particles.
}
\label{spin_charge_diff_Ne}
\end{figure}

In Fig.\ref{spin_charge_diff_size}, we present results on the effect of lattice size used in simulations. The emergence of pFC texture is already clear for $N = 6^2$, $N$ being the number of sites. 
The cluster consists of $7$ sites with their spins almost aligned.
The FM cluster size remains independent of the lattice size $N$, as shown in the figure. Therefore, the formation of such texture is accurately captured even on small lattices.

Another very important issue is to verify that such individual texture survive when the number of electrons increases.
In order to test this explicitly, we perform simulations for different number of electrons on a $12^2$ lattice.
The results are shown in Fig.\ref{spin_charge_diff_Ne}. Clearly, a pFC exists as an isolated texture and number of pFC increases upon increasing
the number of itinerant electrons. It is important to note that a larger uniform-density FM region hosting all the electrons does not appear as the number
of electrons is increased. This suggests that at low electronic density, where the mean electron-electron distance is large, such textures can be realized.

%The only unbiased method to study this model is the hybrid scheme involving exact-diagonalization (ED) of the fermion 
%problem and Monte Carlo (MC) for classical core spins. 
%The difficulty in handling this class of models and details on ED+MC method has been well 
%documented in the context of manganites \cite{Dagotto_book}. In order to achieve larger lattice sizes, which are essential for computing transport 
%properties, we employ the traveling cluster approximation (TCA) \cite{TCA}. This method has proved very  succesSFul in studying DE models on square 
%and cubic lattices relevant for manganites \cite{TCA,SK3}. Since the TCA does not rely on the lattice geometry, we extend the framework to triangular lattices.  
%All of the results in this work are obtained on lattices with $N=12^2$ sites. Typically $\sim 10^4$ MC steps are used for 
%equilibration and a similar number of steps for computing thermal averages on classical spin variables. For electronic properties, which still 
%requires the ED of the full Hamiltonian, $\sim 10^3$ steps are used.